\documentclass[prl,twocolumn,nofootinbib,showpacs,floatfix,superscriptaddress]{revtex4}

\usepackage{amsmath}
\usepackage{graphicx}
\usepackage{bbm}
\usepackage{bm}
\usepackage{latexsym}
\usepackage{amssymb}
\usepackage{booktabs}
\usepackage[ansinew]{inputenc}
\usepackage{rotating}
\usepackage{euscript}
\usepackage{gensymb} 
\usepackage{slashed}
\usepackage{array}

\newcommand{\dd}{{\rm d}}

\begin{document}

\title{Soft-Gluon Resummation and the 
Valence Parton Distribution Function of the Pion}

\author{Matthias Aicher}
\author{Andreas Sch\"afer}
\affiliation{Institute for Theoretical Physics, University of
Regensburg, D-93040 Regensburg, Germany}

\author{Werner Vogelsang}
\affiliation{Institute for Theoretical Physics,
                Universit\"{a}t T\"{u}bingen,
                Auf der Morgenstelle 14,
                D-72076 T\"{u}bingen, Germany}

\date{\today}

      \begin{abstract}
       We determine the valence parton distribution function 
of the pion by performing a new analysis of data for the Drell-Yan 
process $\pi^- N\to \mu^+\mu^-X$. Compared to previous analyses, we
include next-to-leading-logarithmic threshold resummation effects 
in the calculation of the Drell-Yan cross section. As a result of 
these, we find a considerably softer valence distribution at high
momentum fractions $x$ than obtained in previous next-to-leading-order 
analyses, in line with expectations based on perturbative-QCD 
counting rules or Dyson-Schwinger equations.
      \end{abstract}
\maketitle

Although the pion is one of the most important particles in
strong-interaction physics, our knowledge about its internal quark 
and gluon ``partonic'' structure is still rather poor. 
Most of the available information comes from Drell-Yan dimuon
production by charged pions incident on nuclear fixed targets 
\cite{Conway:1989fs,Bordalo:1987cr,DYreview}. These 
data primarily constrain the valence distribution 
$v^\pi\equiv u_v^{\pi^+}= \bar{d}_v^{\pi^+}= d_v^{\pi^-}= \bar{u}_v^{\pi^-}$.
Several next-to-leading order (NLO) analyses of the Drell-Yan 
data have been performed \cite{Sutton:1991ay,Gluck:1999xe,Wijesooriya:2005ir}.
A striking feature has been that the resulting valence
distribution $v^\pi(x,Q^2)$ turned out to be rather hard at high 
momentum fraction $x$, typically showing only a linear $(1-x)^1$
or slightly faster falloff. This finding is at variance with predictions
based on perturbative QCD \cite{Farrar:1979aw},
and calculations using Dyson-Schwinger equations \cite{Hecht:2000xa},
for which the falloff is expected to be $\sim (1-x)^2$. On the other hand,
Nambu-Jona-Lasinio \cite{Shigetani:1993dx} 
and constituent quark models \cite{Frederico:1994dx},
as well as duality arguments \cite{Melnitchouk:2002gh},
favor a linear behavior. The high-$x$ behavior of $v^\pi$ is widely regarded to be an 
important so-far-unresolved problem in strong-interaction 
physics \cite{Holt:2010vj}.

In the kinematic regimes accessed by the fixed-target
Drell-Yan data, perturbative corrections beyond NLO may be
significant~\cite{Shimizu:2005fp}. 
The relation $z=Q^2/x_1 x_2S=1$ sets a threshold 
for the partonic reaction, where $Q$ and $\sqrt{S}$ denote 
the invariant mass of the lepton pair and the overall 
hadronic center-of-mass (c.m.) energy, respectively, and $x_1$ and $x_2$ 
are the momentum fractions of the partons participating
in the hard-scattering reaction. As $z$ increases toward 
unity, little phase space for real-gluon radiation 
remains, since most of the initial partonic energy 
is used to produce the virtual photon. The infrared cancellations
between virtual and real-emission diagrams then leave behind 
large logarithmic higher-order corrections to the
cross sections, the so-called threshold logarithms. These 
logarithms become particularly important in the fixed-target
regime, because here the ratio $Q^2/S$ is relatively large. 
It then becomes necessary to resum the large corrections to all
orders in the strong coupling, a technique known as threshold
resummation. QCD threshold resummation for the Drell-Yan process
has been derived a long time ago~\cite{Sterman:1986aj}. 
It turns out that the threshold logarithms lead to a strong 
increase of the cross section near threshold. Therefore, 
if threshold resummation effects are included, it is possible that 
a much softer valence distribution of the pion is sufficient to describe 
the experimental data. Indeed, as was observed in 
Ref.~\cite{Wijesooriya:2005ir}, the extracted $v^\pi$
already becomes softer when going from the lowest order to NLO, 
where the threshold logarithms first appear.
In this Letter, we will address the impact of resummation effects
on the pion's valence distribution. We will find that indeed a 
falloff $v^\pi\sim (1-x)^2$ even at a relatively low resolution scale 
is well consistent with the Drell-Yan data. We note that the
effects of resummation on parton distributions were also examined
in the context of deep-inelastic lepton scattering~\cite{Corcella:2005us}.

We consider the inclusive cross section for the 
production of a $\mu^+\mu^-$ pair of invariant mass $Q$ 
and rapidity $\eta$ in the process $\pi^{-}(P_1) A(P_2) \to
\mu^+  \mu^-  X$, where $A$ denotes a nucleon or 
nuclear target and $P_1$ and $P_2$ are 
the four-momenta of the initial-state particles. 
According to the factorization theorem, the cross section is written as
\begin{eqnarray}\label{eq:factorization}
\frac{\dd \sigma}{\dd Q^2 \dd \eta} &=& 
\sigma_0 \sum_{a,b} \int_{x_1^0}^1 \frac{\dd x_1}{x_1} \int_{x_2^0}^1 
\frac{\dd x_2}{x_2} f_a^{\pi} (x_1,\mu^2) f_b^A(x_2, \mu^2) \nonumber \\ 
&\times& 
\omega_{ab}(x_1,x_1^0,x_2,x_2^0,Q/\mu) ,
\end{eqnarray}
where $\sigma_0 = 4 \pi \alpha^2/9 Q^2 S$, with  $S = (P_1 + P_2)^2$, and
where 
\begin{align}
 x_{1,2}^0 =\sqrt{\tau}\, e^{\pm \eta}
\end{align}
with $\tau= Q^2/S$. At lowest order, one
has $x_{1,2}=x_{1,2}^0$. 
The sum in Eq.~(\ref{eq:factorization}) 
runs over all partonic channels, with $f_a^{\pi}$ and $f_b^A$ 
the corresponding parton distribution functions
of the pion and the nucleus 
and $\omega_{ab}$ the hard-scattering function. The latter 
can be computed in perturbation theory as a series in the strong 
coupling constant $\alpha_s$, starting from the lowest-order process
$q\bar{q}\to\gamma^*\to\mu^+\mu^-$. 
The functions in~(\ref{eq:factorization}) depend
on the factorization and renormalization scales, which we choose to 
be equal here and collectively denote as $\mu$. 

As discussed above, our goal is to resum large 
logarithmic contributions to $\omega_{q\bar{q}}$ that arise near
partonic threshold $z=1$, where $z=Q^2/\hat{s}=\tau/x_1 x_2$, with
$\sqrt{\hat s}$ the partonic c.m. energy. Resummation may 
be achieved in Mellin moment space, where phase space integrals for
multiple-soft-gluon emission decouple. For the rapidity-dependent cross 
section, it is convenient to also apply a 
Fourier transform in $\eta$ \cite{Sterman:2000pt,Mukherjee:2006uu}. 
Under combined Fourier and Mellin transforms of the cross section,
\begin{equation}
  \sigma(N,M) \equiv \int_0^1 \dd \tau \tau^{N-1} 
\int_{-\ln\frac{1}{\sqrt{\tau}}}^{\ln\frac{1}{\sqrt{\tau}}} 
\dd \eta e^{iM\eta} \frac{\dd \sigma}{\dd Q^2 \dd \eta},
\end{equation}
the convolution integrals in (\ref{eq:factorization}) 
decouple into ordinary products \cite{Mukherjee:2006uu, Sterman:2000pt}.
Defining the moments of the parton distribution functions,
\begin{equation}\label{eq:mellin_moments}
 f^N(\mu^2) \equiv \int_0^1 \dd x x^{N-1} f(x,\mu^2),
\end{equation}
and introducing the corresponding double transform 
of the partonic hard-scattering cross sections,
\begin{equation}
 \tilde{\omega}_{ab}(N,M)\equiv
\int_0^1 \dd z z^{N-1} 
\int_{-\ln\frac{1}{\sqrt{z}}}^{\ln\frac{1}{\sqrt{z}}} 
\dd \hat\eta e^{iM\hat\eta} \omega_{ab},
\end{equation}
where $\hat\eta = \eta - \frac{1}{2}\ln(x_1/x_2)$ is the 
partonic c.m. rapidity, one finds
\begin{equation}\label{eq:sigmaNM}
\sigma(N,M) = \sigma_0 \sum_{a,b} f_a^{\pi, N+i\frac{M}{2}}
f_b^{A, N - i \frac{M}{2}}\tilde{\omega}_{ab}(N,M).
\end{equation}
As was discussed in Refs. \cite{Laenen:1992ey,Mukherjee:2006uu,Bolzoni:2006ky}, 
in the near-threshold limit $z \to 1$ or $N\to\infty$, the 
dependence of $\tilde{\omega}_{ab=q\bar{q}}(N,M)$ on $M$ becomes subleading 
and may be neglected. The resummed expression for $\tilde{\omega}_{q\bar{q}}
(N,M)$
then becomes identical to that for the total (rapidity-integrated) 
Drell-Yan cross section and is given in the $\overline{\mathrm{MS}}$ 
scheme by
\begin{eqnarray}\label{resummed}
 \ln\tilde{\omega}_{q\bar{q}}& =&
C_q\left(\frac{Q^2}{\mu^2},\alpha_s(\mu^2)\right) 
+ 2 \int_0^1 {\rm d}\zeta \frac{\zeta^{N-1} -1}{1-\zeta} \nonumber \\ 
&\times&\int_{\mu^2}^{(1-\zeta)^2 Q^2} \frac{{\rm d}k_\perp^2}{k_\perp^2} 
A_q(\alpha_s(k_\perp)),
\end{eqnarray}
where $A_q(\alpha_s)$ is a perturbative function, whose first two orders
are sufficient for resummation to next-to-leading-logarithmic (NLL) 
order~\cite{Sterman:1986aj}:
\begin{equation}\label{exp_A}
 A_q(\alpha_s) = \frac{\alpha_s}{\pi} A_q^{(1)} + \left(\frac{\alpha_s}{\pi}\right)^2 A_q^{(2)}+ \dots,
\end{equation}
with \cite{Kodaira:1981nh}
\begin{equation}
 A_q^{(1)} = C_F, \quad A_q^{(2)} = \frac{1}{2} C_F\left[C_A\left(\frac{67}{18}-\frac{\pi^2}{6}\right) - \frac{5}{9} N_f\right].
\end{equation}
Here $C_F=4/3$ and $C_A=3$.
The first term in Eq.~(\ref{resummed}) does not originate from  
soft-gluon emission but instead mostly contains hard virtual 
corrections. It is also a perturbative series in $\alpha_s$, and 
we need only its first-order term:
\begin{equation}
C_q = \frac{\alpha_s}{\pi} C_F \left(-4 + \frac{2\pi^2}{3} + \frac{3}{2} \ln \frac{Q^2}{\mu^2}\right) +{\cal O}(\alpha_s^2),
\end{equation}
whose exponentiated form is given in Ref. \cite{Eynck:2003fn}.
As was shown in Ref. \cite{Mukherjee:2006uu},
rapidity dependence is slightly more faithfully
reproduced if one shifts the Mellin moments to $N\pm i M/2$ in
$\tilde{\omega}_{q\bar{q}}$, 
which is a choice that we also adopt here. 

At NLL order, the expression in Eq.~(\ref{resummed}) becomes~\cite{Catani:1996yz} 
\begin{equation}
\ln\tilde{\omega}_{q\bar{q}}=
C_q  + 2  h^{(1)}(\lambda) \ln \bar N 
+ 2 h^{(2)} \left(\lambda, \frac{Q^2}{\mu^2}\right), 
\end{equation}
where $\bar N = N e^{\gamma_E}$ with the Euler constant $\gamma_E$, and
\begin{equation}
\quad \lambda = b_0 \alpha_s(\mu^2) \ln \bar N.
\end{equation}
 The functions $h^{(1)}$ and $h^{(2)}$ collect all leading-logarithmic
and NLL terms in the exponent, which are of the form 
$\alpha_s^k \ln^{k+1} \bar{N}$ 
and $\alpha_s^k \ln^{k} \bar{N}$, respectively. They read
\begin{eqnarray}
h^{(1)}(\lambda) &=& \frac{A_q^{(1)}}{2 \pi b_0 \lambda}\left[2\lambda + 
(1- 2\lambda) \ln(1-2\lambda)\right],\nonumber \\
h^{(2)}(\lambda) &= & \left( \pi A_q^{(1)}b_1
-b_0A_q^{(2)}\right) \frac{2\lambda + \ln (1-2\lambda)}
{2\pi^2 b_0^3} \\
&+&\frac{A_q^{(1)}b_1}{4\pi b_0^3} \ln^2(1-2\lambda)
 + \frac{A_q^{(1)}}{2\pi b_0} \ln(1-2\lambda) \ln\frac{Q^2}{\mu^2},\nonumber
\end{eqnarray}
where
\begin{eqnarray}
b_0 &=& \frac{1}{12\pi}\left(11 C_A - 2 N_f\right) \\
b_1 &=& \frac{1}{24 \pi^2}\left(17 C_A^2 - 5 C_A N_f -3 C_F N_f\right).
\end{eqnarray}

The resummed hadronic rapidity-dependent cross section is obtained 
by taking the inverse Mellin and Fourier transforms of 
Eq.~(\ref{eq:sigmaNM}):
\begin{equation}\label{eq:inverse}
   \frac{\dd \sigma}{\dd Q^2 \dd \eta} = \int_{-\infty}^{\infty} 
\frac{\dd M}{2\pi} e^{-iM\eta} \int_{C-i\infty}^{C+i\infty} 
\frac{\dd N}{2\pi i}\tau^{-N} \sigma(N,M).
\end{equation}
When performing the inverse Mellin transform, the parameter 
$C$ usually has to be chosen in such a way that all singularities 
of the integrand lie to the left of the integration contour. The 
resummed cross section has a Landau singularity at $\lambda = 1/2$
or $\bar{N}=\exp(1/2\alpha_s b_0)$,
as a result of the divergence of the running coupling $\alpha_s$ 
in~(\ref{resummed}) for $k_\perp \to \Lambda_{\mathrm{QCD}}$. 
In the Mellin inversion, we adopt the \textit{minimal prescription} 
developed in Ref. \cite{Catani:1996yz} to deal with the Landau pole, for
which the contour is chosen to lie to the left of the Landau singularity.
An alternative possibility is to perform the resummation directly in 
$z$-space \cite{Becher:2007ty}.
We match the resummed cross section to the NLO one by subtracting 
the $O(\alpha_s)$ expansion of the resummed expression and adding 
the full NLO cross section \cite{Mukherjee:2006uu}. 

The fixed-target pion Drell-Yan data 
\cite{Conway:1989fs,Bordalo:1987cr} are in a kinematic
regime where the partons' momentum fractions are relatively large,
($x \gtrsim 0.3$), and hence the valence quark contributions strongly dominate.
We can therefore only hope to determine the pion's valence distribution
$v^\pi\equiv u_v^{\pi^+}= \bar{d}_v^{\pi^+}= d_v^{\pi^-}= \bar{u}_v^{\pi^-}$.
Following the NLO Gl\"{u}ck-Reya-Schienbein (GRS) analysis
\cite{Gluck:1999xe}, we choose the initial scale $Q_0 = 0.63 
\mbox{ GeV}$ for the evolution and parameterize the valence 
distribution function as
\begin{equation}\label{eq:valence_input}
xv^{\pi}(x,Q_0^2) = N_v x^\alpha (1-x)^\beta (1+\gamma x^\delta),
\end{equation}
subject to the constraint $\int_0^1 v^{\pi}(x,Q_0^2) \dd x = 1$.
Since there is no sensitivity to the 
sea quark and gluon distributions, we adopt them
from the GRS analysis, 
except that we modify the overall normalization of the sea quark
distribution so that the momentum sum rule $\sum_{i=q,\bar{q}g}\int_0^1 \dd x
xf_i(x)=1$ is maintained when we determine the valence distribution.
All distributions are then evolved at NLO to the relevant factorization 
scale $\mu=Q$. 

\begin{table*}[ht]
\caption{Results for our NLL threshold-resummed fits 
to the Fermilab E615 Drell-Yan data \cite{Conway:1989fs}.}
\begin{tabular*}{0.75\textwidth}{@{\extracolsep{\fill}} c  c  c  c  c  c c  }
\hline
 Fit & $2\langle xv^\pi \rangle$ & $\alpha$ & $\beta$ & $\gamma$ & 
$K$ & $\chi^2$ (no. of points) \\
\hline
 1 & 0.55 & $0.15 \pm 0.04 $ & $1.75 \pm 0.04$ & $89.4$   & 
$ 0.999 \pm 0.011$ & $82.8$ $(70)$ \\
 2 & 0.60 & $0.44 \pm 0.07 $ & $1.93 \pm 0.03$ &  $25.5$  & 
$ 0.968 \pm 0.011$ & $80.9$ $(70)$ \\
 3 & 0.65 & $0.70 \pm 0.07 $ & $2.03 \pm 0.06$ & $13.8$   & 
$ 0.919 \pm 0.009$ & $80.1$ $(70)$ \\
 4 & 0.7 & $1.06 \pm 0.05 $ & $2.12 \pm 0.06$ &  $6.7$  & 
$ 0.868 \pm 0.009$ & $81.0$ $(70)$\\
\hline
\end{tabular*}
\label{table:E615_fit}
\end{table*}

The free parameters in Eq.~(\ref{eq:valence_input}) are 
determined by a fit to the pion Drell-Yan data from the Fermilab 
E615 experiment \cite{Conway:1989fs}, applying
threshold resummation as detailed in the previous section. 
The E615 data were obtained by using a 252~GeV $\pi^-$ beam on a tungsten target.
We take into account the nuclear effects in this heavy target by 
using the nuclear parton distribution functions 
from Ref. \cite{deFlorian:2003qf}. We use data points with lepton pair 
mass 4.03 GeV $\leq$ $Q$ $\leq$ 8.53 GeV (between the 
$J/\Psi$ and $\Upsilon$ resonances) and $0<x_F<0.8$. Here, $x_F$ is 
the Feynman variable. In the near-threshold region, which is
addressed by threshold resummation, we can use lowest-order kinematics to 
determine the relation between $x_F$ and the rapidity $\eta$:
\begin{equation}
 x_F = x_1^0 - x_2^0 = \sqrt{\tau} \sinh(\eta).
\end{equation}
Since the E615 data have a nominal overall systematic 
error of $16\%$,  we introduce a normalization factor $K$ 
that multiplies the theoretical cross section.
We find that the parameter $\delta$ in~(\ref{eq:valence_input}) is not 
well-determined, and we hence fix it to $\delta = 2$, a 
value roughly preferred by the fit. 
In order to obtain a better picture of the physical content of
our determined pion valence distribution, we perform fits for
several different values of its total momentum fraction 
$\langle xv^\pi \rangle = \int_0^1 x v^\pi(x,Q_0^2)$. 
Fixing $\langle xv^\pi\rangle$ makes one parameter in 
Eq.~(\ref{eq:valence_input}) redundant, which we choose to be $\gamma$.
We hence fit the remaining three free parameters $\alpha$, $\beta$ and $K$ 
to the 70 data points using a $\chi^2$ minimization procedure.

\begin{figure}[ht]
  \includegraphics[height=0.77\columnwidth, angle = 90]{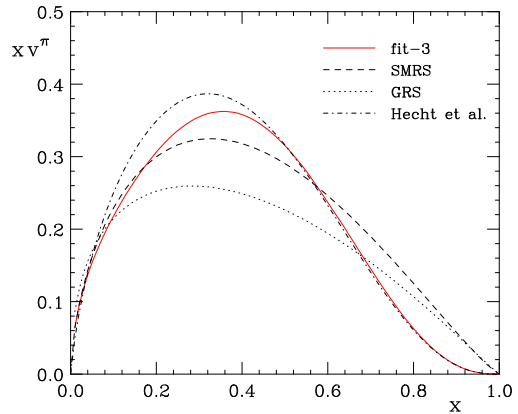}
  \caption{The pionic valence ($v^\pi$) distribution obtained 
from our fit 3 to the E615 Drell-Yan data at $Q = 4 \mbox{ GeV}$, 
compared to the NLO parameterizations of \cite{Sutton:1991ay} (Sutton-Martin-Roberts-Stirling) 
and \cite{Gluck:1999xe} (GRS) and to the distribution obtained
from Dyson-Schwinger equations \cite{Hecht:2000xa}.}
  \label{fig:pdfs}
\end{figure}

The results are shown in Table \ref{table:E615_fit},
for four different values of the total valence quark momentum fraction
$2\langle xv^\pi \rangle$. One observes that fit 3 for which 
the valence carries $65\%$ of the pion's momentum is preferred, 
with slightly higher or lower values also well acceptable. 
Most importantly, all fits show a clear preference for a falloff
much softer than linear, with fits 2,~3, and 4 having a value of 
$\beta$ very close to $2$. This is the central result of our work. 
The valence distribution $xv^\pi$ for our best fit 3 
is shown in Fig.~\ref{fig:pdfs}, evolved to 
$Q = 4 \mbox{ GeV}$. At this scale it behaves
as $(1-x)^{2.34}$. Valence distributions obtained from previous NLO analyses 
\cite{Sutton:1991ay, Gluck:1999xe}, which have a roughly linear 
behavior at high $x$, and from calculations using 
Dyson-Schwinger equations \cite{Hecht:2000xa}, for which 
$v^\pi\sim (1-x)^{2.4}$, are also shown. 
We note that for all our fits the factors $K$ lie well within the 
normalization uncertainty of the data. 
\begin{figure}[ht]
  \includegraphics[height=0.77\columnwidth, angle = 90]{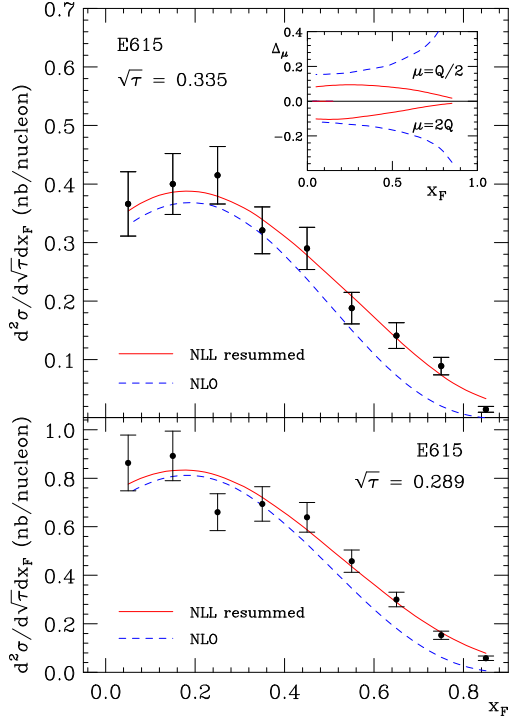}
  \caption{Comparison of our NLL-resummed Drell-Yan cross section 
based on fit 3 to some of the E615 Drell-Yan data. The inset in the upper
figure shows the scale variation of the resummed and the NLO cross sections
(see the text).}
  \label{fig:E615}
\end{figure}

\begin{figure}[t!]
  \includegraphics[height=0.77\columnwidth, angle = 90]{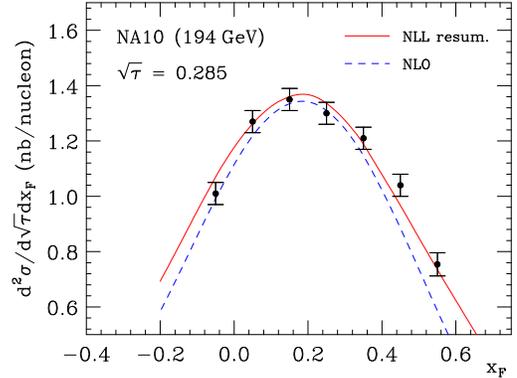}
  \caption{Comparison of our resummed results for fit 3 to some of 
the CERN NA10 $\pi^-$-$W$ data \cite{Bordalo:1987cr} at 
pion beam energy 194 GeV. We have used a 
normalization factor $K = 1.045$.}
  \label{fig:NA10}
\end{figure}

In Fig.~\ref{fig:E615} we compare the resummed Drell-Yan cross section
obtained for fit 3 to some of the E615 data. We have chosen 
the factorization and renormalization scale $\mu=Q$. As one can see from 
the figure and from Table \ref{table:E615_fit}, the data are very 
well described. This also holds true for the CERN 
NA10 \cite{Bordalo:1987cr} Drell-Yan data, which
were not included in our fit, and to which we compare in Fig.~\ref{fig:NA10}.
We also show the results obtained for our fit 3 when
using only  NLO (i.e., unresummed) partonic cross 
sections in the calculation.
As seen in Fig.~\ref{fig:E615}, these fall off too rapidly at large $x_F$. 
The inset in Fig.~\ref{fig:E615} shows the variation of our NLO
and resummed cross sections with $\mu$, in terms of the quantity
$\Delta_\mu\equiv \left[\sigma(\mu')-\sigma(\mu=Q)\right]/\sigma(\mu=Q)$, with 
$\mu'=Q/2$,~$2Q$. As one can see, the scale uncertainty is significantly
reduced after resummation and becomes smaller than the experimental
uncertainties, in particular in the region of high $x_F$. This implies
that our findings for the pion's valence distribution are stable
with respect to the main theoretical uncertainty in the calculation.

In conclusion, we have determined a new valence parton distribution function 
for the pion by reanalyzing pion-nucleon Drell-Yan data 
including threshold-resummed contributions to the cross section. 
The obtained valence distribution is much softer in the high-$x$ 
regime than that found in an NLO analysis, behaving roughly 
as $(1-x)^2$, in agreement with predictions from perturbative QCD 
and nonperturbative Dyson-Schwinger equation approaches. 

We are grateful to F.\ Yuan for useful communications. 
This work was supported by BMBF. M.A. was supported by a grant of 
the ``Bayerische Elitef\"orderung.''
W.V.'s work has been supported by the U.S. Department of Energy 
(Contract No. DE-AC02-98CH10886) and by the 
Alexander von Humboldt Foundation.

\begin{thebibliography}{99}

\bibitem{Conway:1989fs}
  J.~S.~Conway {\it et al.}  [E615 Collaboration],
  Phys.\ Rev.\  D {\bf 39}, 92 (1989).

\bibitem{Bordalo:1987cr}
  P.~Bordalo {\it et al.}  [NA10 Collaboration],
  Phys.\ Lett.\  B {\bf 193}, 373 (1987);
  B.~Betev {\it et al.}  [NA10 Collaboration],
  Z.\ Phys.\  C {\bf 28}, 9 (1985).

\bibitem{DYreview} For a review, see: W.~J.~Stirling and M.~R.~Whalley,
  J.\ Phys.\ G {\bf 19}, D1 (1993).


\bibitem{Sutton:1991ay}
  P.~J.~Sutton {\it et al.},
  Phys.\ Rev.\  D {\bf 45}, 2349 (1992).

\bibitem{Gluck:1999xe}
  M.~Gl\"{u}ck, E.~Reya and I.~Schienbein,
  Eur.\ Phys.\ J.\  C {\bf 10}, 313 (1999).

\bibitem{Wijesooriya:2005ir}
  K.~Wijesooriya, P.~E.~Reimer and R.~J.~Holt,
  Phys.\ Rev.\  C {\bf 72}, 065203 (2005).

\bibitem{Farrar:1979aw}
   G.~R.~Farrar and D.~R.~Jackson,
   Phys.\ Rev.\ Lett.\  {\bf 43}, 246 (1979);
   E.~L.~Berger and S.~J.~Brodsky,
   Phys.\ Rev.\ Lett.\  {\bf 42}, 940 (1979);
   S.~J.~Brodsky and F.~Yuan,
   Phys.\ Rev.\  D {\bf 74}, 094018 (2006);
   F.~Yuan,
   Phys.\ Rev.\  D {\bf 69}, 051501 (2004).


 \bibitem{Hecht:2000xa}
   M.~B.~Hecht, C.~D.~Roberts and S.~M.~Schmidt,
   Phys.\ Rev.\  C {\bf 63}, 025213 (2001).

 \bibitem{Shigetani:1993dx}
   T.~Shigetani, K.~Suzuki and H.~Toki,
   Phys.\ Lett.\  B {\bf 308}, 383 (1993).

 \bibitem{Frederico:1994dx}
   T.~Frederico and G.~A.~Miller,
   Phys.\ Rev.\  D {\bf 50}, 210 (1994);
   A.~Szczepaniak, C.~R.~Ji and S.~R.~Cotanch,
   Phys.\ Rev.\  D {\bf 49}, 3466 (1994).

 \bibitem{Melnitchouk:2002gh}
   W.~Melnitchouk,
   Eur.\ Phys.\ J.\  A {\bf 17}, 223 (2003).



\bibitem{Holt:2010vj}
  R.~J.~Holt and C.~D.~Roberts,
  Rev.\ Mod.\ Phys.\  {\bf 82}, 2991 (2010).


\bibitem{Shimizu:2005fp}
  H.~Shimizu {\it et al.},
  Phys.\ Rev.\  D {\bf 71}, 114007 (2005).

\bibitem{Sterman:1986aj}
  G.~F.~Sterman,
  Nucl.\ Phys.\  B {\bf 281}, 310 (1987);
  S.~Catani and L.~Trentadue,
  Nucl.\ Phys.\  B {\bf 327}, 323 (1989).

\bibitem{Corcella:2005us}
  G.~Corcella and L.~Magnea,
  Phys.\ Rev.\  D {\bf 72}, 074017 (2005).

\bibitem{Sterman:2000pt}
  G.~F.~Sterman and W.~Vogelsang,
  JHEP {\bf 0102}, 016 (2001).

\bibitem{Mukherjee:2006uu}
  A.~Mukherjee and W.~Vogelsang,
  Phys.\ Rev.\  D {\bf 73}, 074005 (2006).

\bibitem{Laenen:1992ey}
  E.~Laenen and G.~Sterman, FERMILAB Report No. CONF-92-359-T (unpublished).

\bibitem{Bolzoni:2006ky}
  P.~Bolzoni,
  Phys.\ Lett.\  B {\bf 643}, 325 (2006).

\bibitem{Kodaira:1981nh}
  J.~Kodaira and L.~Trentadue,
  Phys.\ Lett.\  B {\bf 112}, 66 (1982).

\bibitem{Eynck:2003fn}
  T.~O.~Eynck, E.~Laenen and L.~Magnea,
  JHEP {\bf 0306}, 057 (2003).

\bibitem{Catani:1996yz}
  S.~Catani {\it et al.}, 
  Nucl.\ Phys.\  B {\bf 478}, 273 (1996).

\bibitem{Becher:2007ty}
  T.~Becher, M.~Neubert and G.~Xu,
  JHEP {\bf 0807}, 030 (2008).

\bibitem{deFlorian:2003qf}
  D.~de Florian and R.~Sassot,
  Phys.\ Rev.\  D {\bf 69}, 074028 (2004).


\end{thebibliography}
\end{document}